# Evaluating Compression and Nanoindentation in FCC Nickel: A Methodology for Interatomic Potential Selection

**K. Cichocki** [1], F. J. Dominguez-Gutierrez[2)*], L. Kurpaska[2)], K. Muszka[1)]

[1)] *Faculty of Metals Engineering and Industrial Computer Science, AGH University of Krakow, Al. Mickiewicza 30, 30-059 Krakow, Poland.*
[2)] *National Centre for Nuclear Research, NOMATEN CoE MAB+, Andrzeja Soltana 7, 05-400 Otwock-Swierk, Poland*

## Abstract

We performed molecular dynamics simulations to investigate the mechanical response of face-centered cubic (FCC) nickel under uniaxial compression and nanoindentation using traditional interatomic potentials, including the Embedded Atom Method (EAM) and Modified Embedded Atom Method (MEAM). By calculating the generalized stacking fault energy (GSFE), we analyzed the dissociated slip paths responsible for stacking fault formation and partial Shockley dislocations during mechanical loading. Our findings highlight the critical importance of selecting appropriate interatomic potentials to model compression and nanoindentation tests accurately, aligning simulations with experimental observations. We propose a practical methodology for identifying empirical interatomic potentials suitable for mechanical testing of single-element materials. This approach establishes a benchmark for FCC nickel simulations and provides a basis for extending these methods to more complex Ni-based alloys, facilitating comparisons with experimental results such as those from electron microscopy.

Corresponding author E-mail: javier.dominguez@ncbj.gov.pl

## 1. Introduction

Nickel is a well-known material with FCC structure and is characterized by good mechanical properties like good toughness and ductility. Nickel also is a good conductor of heat and electricity [1], with superior corrosion resistance in caustic or nonoxidizing acidic solutions, and in gaseous halogens [2]. Because of the mentioned properties, nickel is used as an alloying element in stainless steels and in Ni-Cu, Ni-Cr-Fe [3] alloys, for example, to increase strength properties and corrosion resistance. In addition, due to their mechanical properties, over the years a number of nickel-based alloys have been developed which are characterized not only by good corrosion resistance, but also by good creep resistance at high temperatures. Nickel-based alloys [4][5] have found widespread use in industrial applications such as the aerospace industry[6] and in the nuclear power industry [7]. However, one must remember that this material activates under neutron flux, so its high content should be avoided when considering internals that could suffer large neutron exposure. For this reason its content should be balanced in the FCC alloys and one way to do that is to optimize mechanical properties of the material by predicting their deformation behavior by MD simulations. Hence, usage of different available interatomic potentials should be considered at first.

Nanoindentation is a technique for measuring the mechanical properties of materials in small areas. The test involves inserting a suitable indenter of various geometries into the material in a direction perpendicular to its plane, with simultaneous measurement of the force versus displacement of the indenter. This method allows obtaining material parameters such as Young's modulus, hardness or even a yield point via analysis of the stress-strain curves [8][9]. In the case of metals and alloys, the method makes it possible to determine mechanical properties and material parameters in individual grains orientation. Nanoindentation simulations allow more accurate analysis of material behavior, and are used to characterize material behavior during force application. Simulations allow additional analysis of dislocation nucleation [9], the formation of stacking faults [11] and the analysis of deformation mechanisms [12]. A number of publications show that MD simulations of nanoindentation allow a good analysis of the presented phenomena [13-21].

However, MD simulations are not without problems, one of the main ones being the selection of an appropriate potential for a given type of calculation [17-21].

In this work, we conducted molecular dynamics simulations to investigate the mechanical response of face-centered cubic (FCC) nickel under uniaxial compression and nanoindentation using traditional interatomic potentials, including the Embedded Atom Method (EAM) and Modified Embedded Atom Method (MEAM) formulations. By calculating the generalized stacking fault energy (GSFE) for each potential, we analyzed the dissociated slip paths responsible for stacking fault formation and the emergence of partial Shockley dislocations during mechanical loading. Our results emphasize the importance of establishing a benchmark for such simulations by identifying interatomic potentials capable of accurately modeling both compression and nanoindentation tests in a manner consistent with experimental observations. While machine learning-based interatomic potentials are increasingly popular due to their potential for improved accuracy, they often require robust training datasets and significant computational resources for large-scale simulations. In contrast, our study presents a practical methodology for selecting empirical interatomic potentials to perform mechanical tests on single-element materials. This approach provides a foundation for comparison with electron microscopy observations and offers a pathway to extend the simulations to more complex Ni-based alloys.

## 2. Computational Methods

To perform our simulations, we use Large-scale Atomic/Molecular Massively Parallel Simulator (LAMMPS) [22] software which allows us to study the behavior of materials under a wide range of conditions. One of our goals is to accurately model plastic deformation, which is a crucial aspect of how materials respond to external loads. In this study, we employ traditional interatomic potentials based on the Embedded Atom Method (EAM) and Modified Embedded Atom Method (MEAM) [22–25]. The selected potentials include EAM-1 by Malerba et al. [22], which models Fe–Ni systems with stable ordered intermetallic phases $L1_0$-FeNi and $L1_2$-$FeNi_3$, serving as a basis for simulating nickel's mechanical behavior. EAM-2, developed by Stoller et al. [23], focuses on modeling high-energy collision cascades in nickel, based on Density Functional Theory (DFT) calculations [23][24], and effectively captures the mechanical response of irradiated nickel. EAM-3 by Zhou et al. [25] is used to model NiFe alloys, offering detailed information on dislocation dynamics under mechanical testing, which supports benchmarking for nickel-based tests. Additionally, we apply the MEAM potential by Choi et al. [26], designed for the FCC Cantor alloy NiFeCrCoMn, which provides an accurate basis for simulating nickel's mechanical response in complex alloy systems. We exclude recent machine learning-based interatomic potentials from our study due to their computational demands. Our focus on large-scale simulations with modest computational resources limits the feasibility of using these novel potentials, which often require access to supercomputers.

### 2.1 Sample preparation

We initially defined the FCC Ni sample with crystal orientations along [001], [101], and [111] by maintaining a density of approximately 8.78 g/cm³. To achieve energy-optimized structures with an energy tolerance of $10^{-6}$ eV, we employed the FIRE (fast inertial relaxation engine) 2.0 protocol, which effectively optimized the sample's energy and identified the lowest energy structure. Subsequently, the samples underwent a thermalization process at 300 K for 100 ps, utilizing a Nose-Hoover NPT thermostat with a time constant of 100 fs until the sample reaches a homogenous temperature and pressure profile [17-21]. To dissipate any artificial heat, a relaxation period of 10 ps was applied afterwards [19-21].

### 2.2 Uniaxial compression test

MD simulations were performed for compression scenarios using a numerical cell with dimension (8.82,7.0,7.5) nm for [001], (8.82, 9.8, 10.6) nm for [101], and (6.25, 12.1, 18.3) nm for [111] with $8.4 \times 10^4$, $4.48 \times 10^6$ , and $4.48 \times 10^6$ Ni atoms, respectively. Each sample is optimized and equilibrated to 300K as explained before, MD simulations were then executed with a strain rate of $10^9$ $s^{-1}$ on the z-direction, where the atomic positions were remapped at each computational step to the instantaneous dimensions of the simulation cell [13, 21, 33]. This value for the strain rate is chosen due to the computational limitations inherent to MD simulations, which can only model systems on the scale of nanometers and nanoseconds. To observe meaningful deformation phenomena within these constraints, high strain rates are necessary to induce mechanical responses within the short simulation timescales. While this deviates from experimental conditions, the qualitative insights gained often remain valid, particularly for understanding mechanisms at the atomic scale. This approach ensured strict displacement-controlled straining of the cuboidal cell along the stretching direction, while a barostat was applied in the other two principal directions. The applied stress $\sigma_{zz}$ was directly obtained from the MD simulations by accounting for the number of atoms and their atomic volumes. The imposed uniaxial strain was calculated as $\varepsilon = L_z - L_{z0} / L_{z0}$, where $L_z$ is the instantaneous cell length along the straining z-axis, and $L_{z0}$ is the initial cell length prior to compression. A time step of 2 fs was used throughout the simulations [13].

### 2.3 Nanoindentation test

A numerical cell of $4.5 \times 10^6$ , $4.48 \times 10^6$, and $4.7508 \times 10^6$ Ni atoms for the main crystal orientations [001] with a sample size of (52.3,52.3,17.6) nm, [101] with a size of (35.5, 34.35, 45) nm, and [111] with (46.3, 46.3, 30) nm, respectively, are optimized and thermalized to 300 K. Before conducting the nanoindentation test, we divided the prepared sample into three sections along the z direction to establish appropriate boundary conditions. The two lowermost layers were kept frozen, covering approximately 0.02 times the thickness increment (xdz), which maintained the stability of the Ni atoms during the nanoindentation process. Additionally, a thermostatic region, located above the frozen layers, was included to effectively dissipate heat generated during the nanoindentation process. This thermostatic region had a thickness of approximately 0.08 times the thickness increment (xdz). The remaining layers constituted a region with dynamical atoms, where the interaction among atoms occurred as the indenter tip modified the surface structure and morphology. Furthermore, a 5 nm vacuum section was incorporated above the material sample [13-21]. In our simulation setup, we employed a non-atomic repulsive imaginary (RI) rigid sphere as the indenter tip, with a force potential given by the equation $F(t) = K\,(r(t) - R)^2$. The force constant, K = 236 eV/$A^3$, denotes the specified force constant, ensuring high stiffness for our indenter tip and indenter tip radius R = 12 nm. The position of the indenter tip's center, denoted as r(t), varied with time according to the equation $r(t) = (x_0, y_0, z_0 \pm vt)$. Here, $x_0$ and $y_0$ represented the center coordinates on the xy plane of the surface sample, $z_0$ = 0.5 nm indicated the initial gap between the surface and the indenter tip, and the indenter tip moved at a speed of v = 20 m/s with a positive value for the loading process and negative for the unloading one [17-21]. The molecular dynamics (MD) simulations were conducted with periodic boundary conditions applied along the x and y axes to simulate an infinite surface. Each process was carried out for a duration of 225 ps, with a time step of $\Delta t$ = 1 fs. The maximum duration of the simulation was determined by the desired duration of the loading stages. The load-displacement curve is then obtained by plotting the force on the indenter tip as a function of its displacement relative to the surface, as the indenter is driven into the material over time [13-21].

## 3. Results

The generalized stacking fault energy (GSFE) is a valuable surrogate property for predicting the plastic response of a material, including its dislocation and twinning behavior. Thus, the variation of the system energy as a crystal undergoes translational slip along specific directions on a slip plane is known as the γ-surface. The maximum energy point on this surface, $\gamma_{usf}$, corresponds to the unstable stacking fault energy, which represents the energy barrier for dislocation nucleation at stress concentrations. A metastable point on the γ-surface, $\gamma_{sf}$, corresponds to the dislocation dissociation energy. To compute the GSFE, periodic boundary conditions were applied along the cut plane, using a replicated sample from a FCC unit cell with 13957 Ni atoms and a dimension of $4.31 \times 2.59 \times 14.02$ $nm^3$. Displacements are applied in equal increments, each representing 0.1 of the Burgers vector magnitude. Following each displacement, the top and bottom atomic layers are fixed, while the remaining layers relax exclusively in the *y* direction. This relaxation process is performed using energy minimization with the conjugate gradient method and is considered complete when either (i) the ratio of the energy change between successive iterations to the most recent energy magnitude is less than $10^{-12}$, or (ii) the global force vector magnitude for all atoms is less than or equal to $10^{-12}$ eV/Å. Subsequently, the stacking fault energy can be calculated as:

$$\gamma GSFE = \frac{Es - E0}{ASF} \quad (1)$$

where Es represents the energy of the sample at a given displacement, and E0 denotes the energy for the perfect sample, ASF stands for the stacking fault area.

Figure 1 displays the computed GSFE for various displacement vectors, where atomic positions were relaxed only perpendicular to the cut plane showing the γ-lines along the ⟨110⟩/2 in a) and ⟨112⟩/2 in b) directions on the {111} plane (most dense plane in FCC) determined from DFT [31,32] and traditional interatomic potentials. Our results indicate that the MEAM potential overestimates the GSFE values for the ⟨110⟩ slip system relative to DFT calculations [31,32]. However, all the investigated potentials overestimate the GSFE for the ⟨112⟩ direction compared to DFT [31,32]. This discrepancy may explain the formation of unphysical defects observed in classical MD simulations to correctly simulate dislocation nucleation and dislocation dissociation behavior.

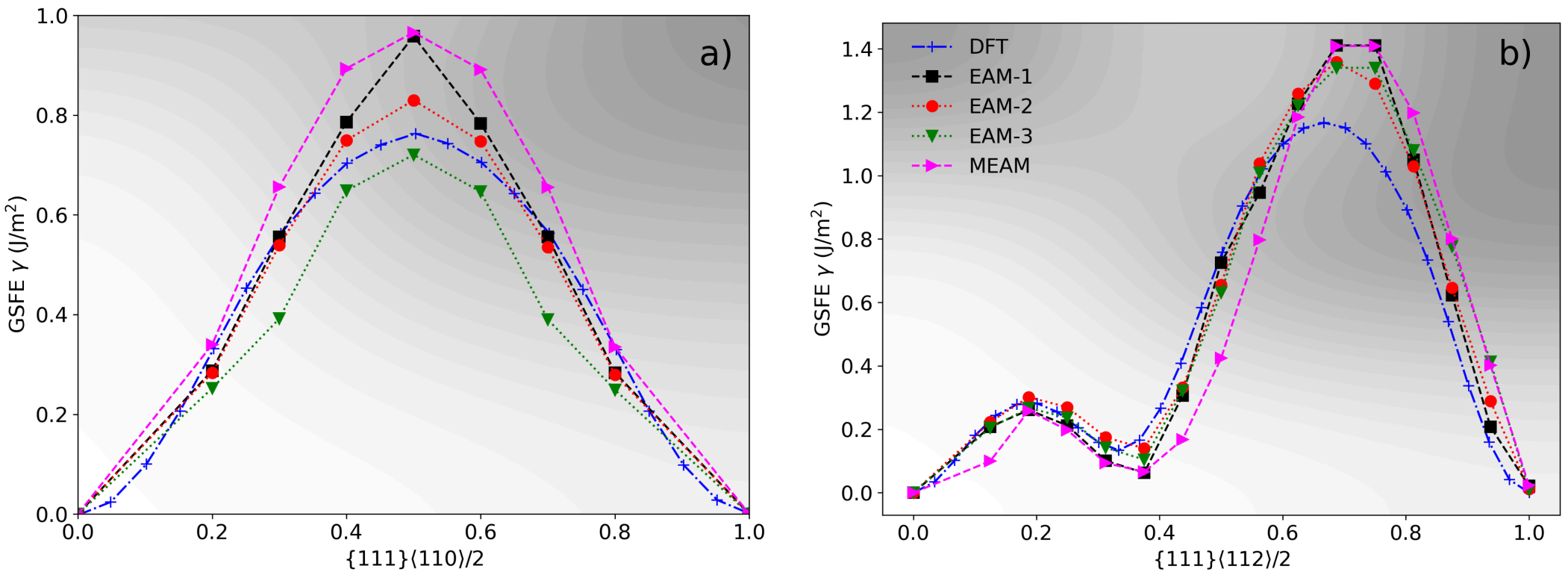

Fig 1. Generalized Stacking Fault Energy (GSFE) surfaces for slip systems ⟨110⟩/2 (a) and ⟨112⟩/2 (b) on the {111} plane, calculated using DFT and various interatomic potentials. Atomic positions were relaxed perpendicular to the cut plane.

Shockley partial dislocations are separated by a stable stacking fault. Accurately modeling the dissociation of dislocations into partial dislocations and the subsequent separation of these partials is essential for precise modeling of plastic deformation mechanisms. In Fig 2, we present results for the GSFE for the *γ*-surface on the loosest packing {111} planes by different interatomic potentials. Our results indicate that the minimum energy path for the dissociation of full dislocations (⟨110⟩/2 or ⟨112⟩/2) into Shockley partial dislocations on the {111} plane aligns with previously reported data [30]. However, the differences in energy values obtained from various computational approaches may lead to significant variations in the predicted behavior of dislocation nucleation and evolution. Specifically, these discrepancies could influence the likelihood of forming different dislocation types, such as Hirth or Frank dislocations, during external mechanical loading. Here, the MEAM potentials are recommended to model the nucleation of Shockely partials during mechanical testing.

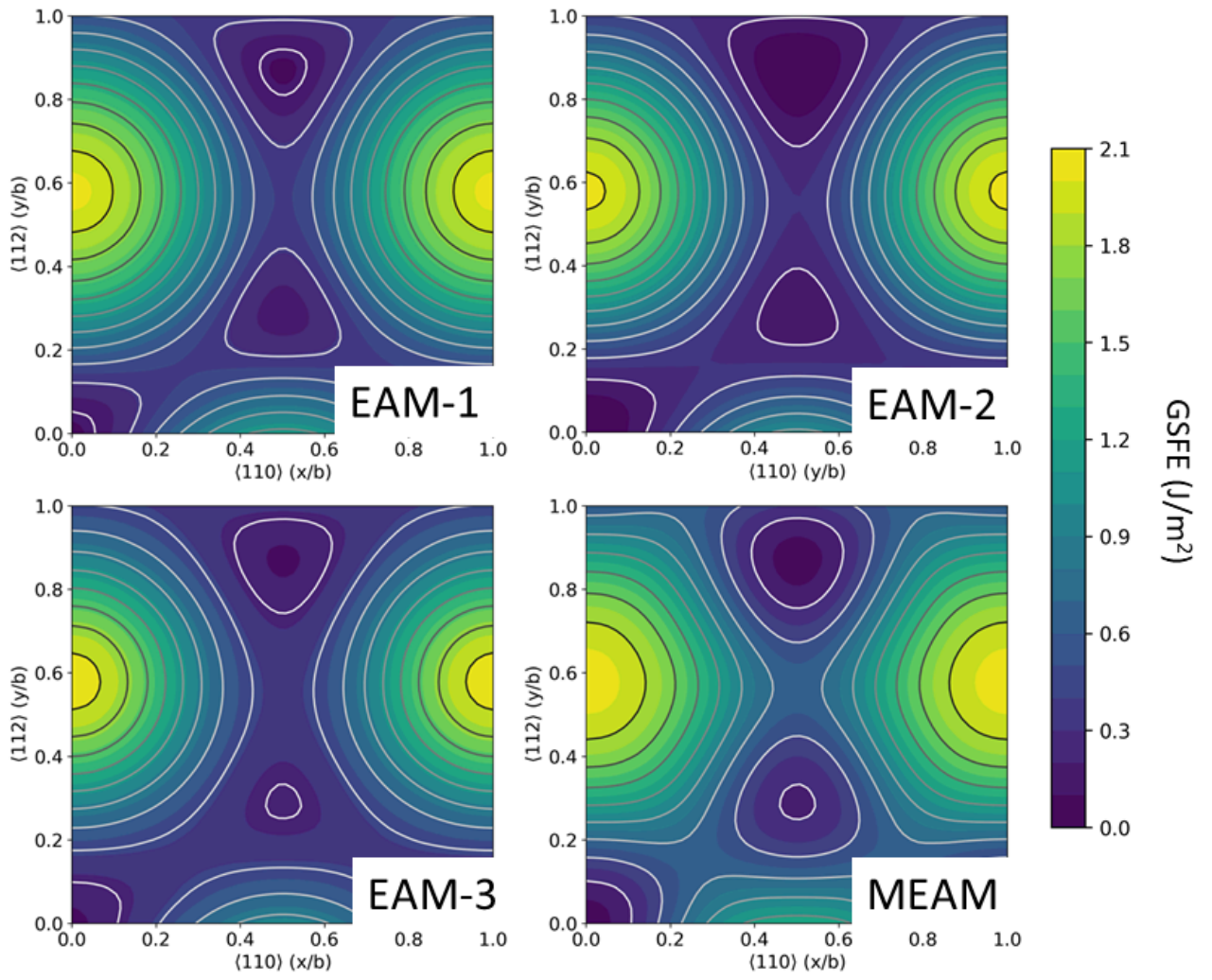

Fig 2. GSFE for the γ-surface on the loosest packing {111} planes by different interatomic potentials showing the difference to model plastic deformation by different approaches.

To validate the suitability of the interatomic potentials for simulations under high-stress conditions, we evaluate the elastic constants over a pressure range relevant to the loading process [13,14]. In the MD approach, the calculation of elastic constants $C_{ij}$ at zero temperature is based on analyzing the stress-strain response of a system. Initially, the material undergoes energy minimization to achieve equilibrium, ensuring the stress tensor accurately represents the system's relaxed state. Small, predefined strains are then incrementally applied to the simulation box, modifying its dimensions and, if necessary, its shape. For each strain increment, the components of the stress tensor are calculated. Pressure plays a crucial role in these calculations, as it directly influences the equilibrium state of the system. At zero pressure, the interatomic spacing and stress tensor are idealized for elastic constant computations, followed by increasing the pressure values during the optimization of the sample before the calculation of the elastic constants $C_{ij}$ . From the stress-strain relationship, the second derivatives of the strain energy are computed to obtain elastic constants, providing a detailed characterization of the material's mechanical response. This methodology ensures the calculated elastic constants reflect intrinsic material properties at the specified thermodynamic conditions. Additionally, we examine elastic stability by applying the spinodal, shear, and Born criteria under hydrostatic pressure, P [14]. These criteria, expressed as $M_1 = C_{11} + 2C_{12} + P > 0$, $M_2 = C_{44} - P > 0$, and $M_3 = C_{11} - C_{12} - 2P > 0$, are illustrated in Figure 3. The materials demonstrate stability from 0 GPa to 50 GPa, which informs the setup of the numerical environment for nanoindentation simulations. However, for the EAM-2 potential, the Born criterion becomes negative beyond 40 GPa, rendering it unsuitable for high-pressure simulations.

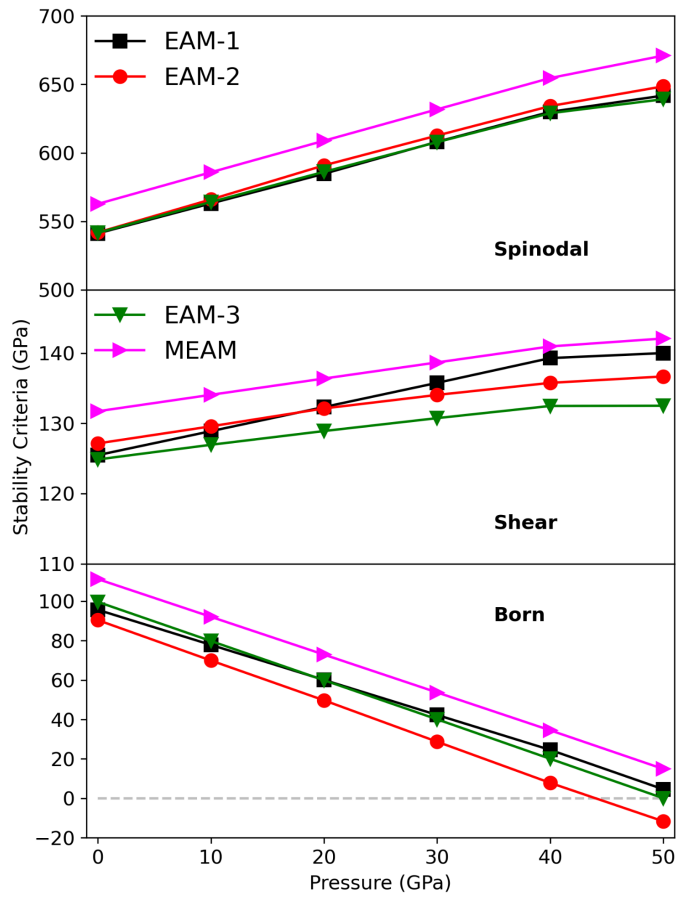

Fig. 3. Spinodal, shear, and Born stability criteria with hydrostatic pressure for different interatomic potentials. The pressure range showing the stability to set up the numerical conditions in the MD simulations.

## 3.1 Uniaxial compression test

In uniaxial compression simulations, we apply compressive strain along the z-axis using different interatomic potentials. This approach reduces the dimension along the compression axis while allowing the other dimensions to either remain constant. A controlled strain rate is applied, as higher strain rates can induce brittle behavior by limiting the time available for atomic rearrangements. In Fig. 4, we present the stress-strain responses for primary crystal orientations [001], [101], and [111] using several interatomic potentials. The results indicate that the mechanical response is similarly captured by the EAM-1 and EAM-3 potentials across different crystal orientations. In contrast, EAM-2 describes the elastic-to-plastic transition more gradually for the [001] orientation and aligns more closely with MEAM results for the [111] orientation, consistent with the GSFE profile. The MEAM potential captures a response more closely resembling typical experimental data for all crystal orientations, as it accounts for a more comprehensive description of nearest-neighbor interactions. Additionally, the ultimate stress is higher in simulations with the MEAM potential than with the EAM potentials, indicating that the initiation of plastic deformation requires greater strain with MEAM, which may better reflect experimental observations.

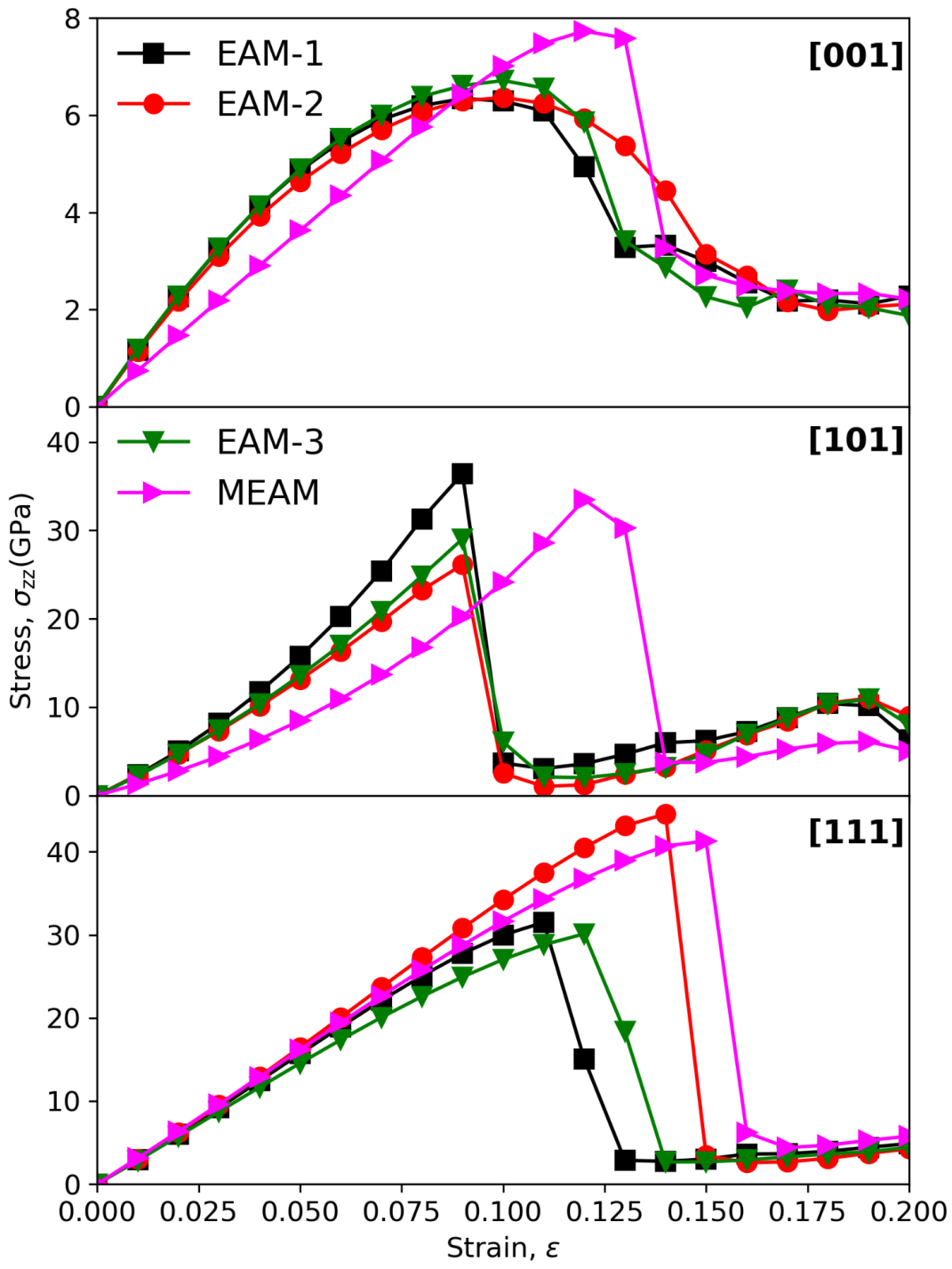

Fig 4. Stress-strain curve for uniaxial compression simulations for the main crystal orientations [001], [101], and [111] by using different interatomic potentials.

In Fig. 5, we present results for the dislocation length as a function of the strain for the [001] crystal orientation by different interatomic potentials. This analysis was performed using the Dislocation Extraction Algorithm (DXA) [29], which identifies and categorizes dislocation structures within atomistic microstructures. Dislocations were classified by their Burgers vectors into several types: 1/2⟨110⟩ (perfect), 1/6⟨112⟩(Shockley), 1/6⟨110⟩ (stair-rod), 1/3⟨100⟩ (Hirth), and 1/3⟨111⟩ (Frank). We observed that the nucleation of partial 1/6⟨112⟩ Shockley dislocations dominated during the tensile test at a strain of 0.12 and 0.14 respect to the used interatomic potentials, which is characteristic of FCC structures. We also observed that different interatomic potentials model the nucleation of Frank dislocations in distinct ways, likely due to variations in their representation of Peierls–Nabarro energies. In contrast, the nucleation of Shockley and Hirth dislocations is modeled similarly across the different potentials, with only minor differences in the strain required to initiate plastic deformation.

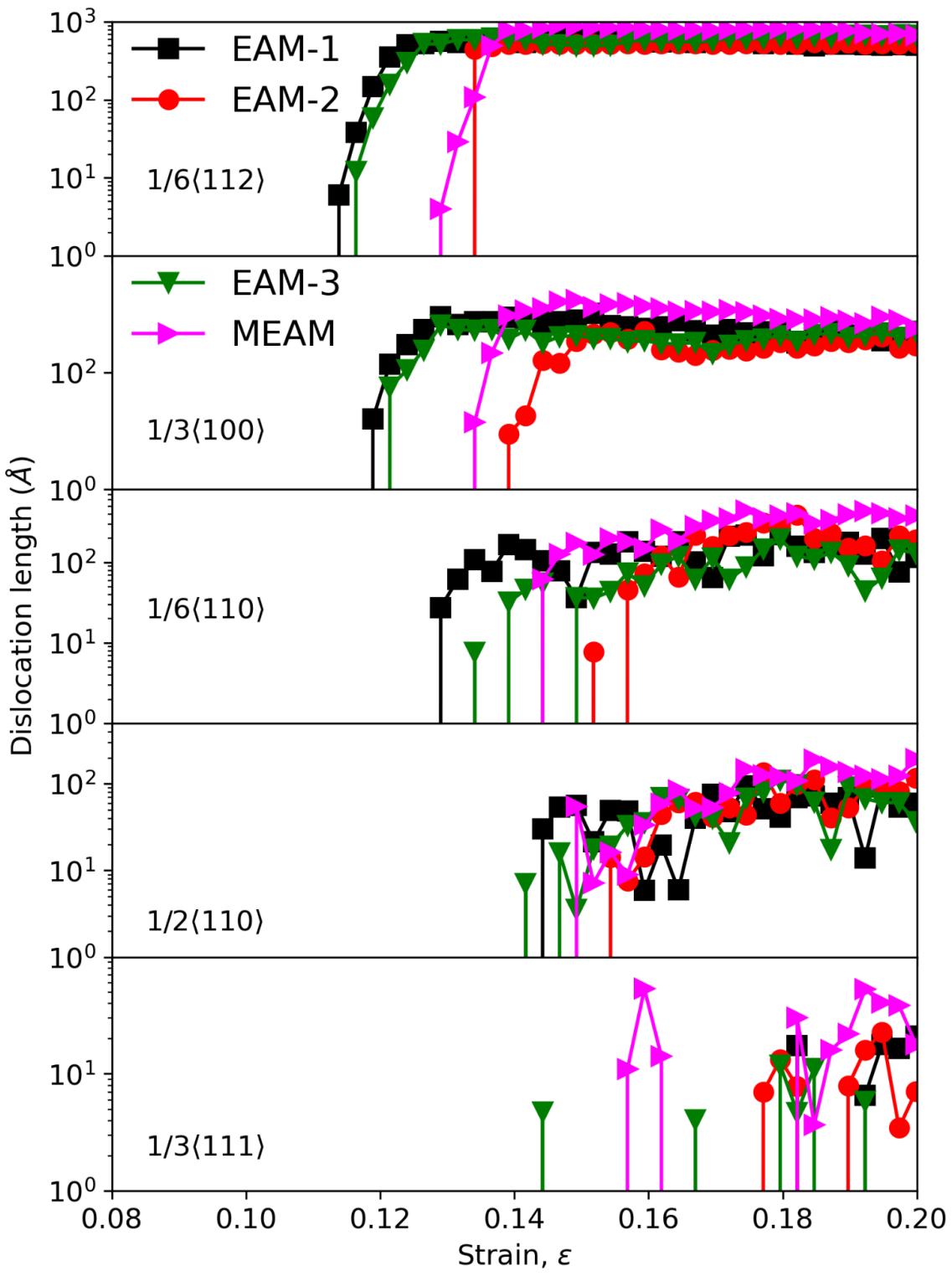

Fig. 5. Dislocation length as a function of the applied strain for [001] Ni by different interatomic potentials.

In a compression test, the mechanical response of a material is largely governed by atomic-scale processes such as dislocation nucleation and stacking fault formation, and it can be observed at the initiation of the plastic deformation of the materials. At the onset of mechanical loading, the material deforms elastically, with atomic rearranging, as observed in Fig 4 for different crystal orientation and modeled in similar way by the chosen interatomic potential. This stage produces a linear stress-strain relationship, and the material will revert to its original shape if the load is removed. As the applied stress increases and reaches the yield strength, plastic deformation begins. Dislocations start to nucleate, as shown in Fig 5. These initial dislocations allow for permanent deformation, as they can move and multiply under load. In FCC materials, partial dislocations move and can disrupt the atomic stacking sequence (e.g., ABCABC in FCC materials), leading to the formation of stacking faults. The ease of stacking fault formation depends on the material's stacking fault energy (SFE), as shown in Fig 2; low-SFE materials favor wider stacking faults as partial dislocations separate more readily. These faults are often created by Shockley partial dislocations, which displace atoms in a way that results in intrinsic stacking faults. In FCC metals with very low SFE, extensive stacking fault formation can even lead to twinning, where portions of the crystal form mirror-image atomic arrangements relative to the main lattice.

As compression deformation continues, dislocations multiply and interact with stacking faults, leading to increasingly complex behavior. Dislocation interactions can lead to obstacles such as dislocation pinning, cross-slip, or the formation of junctions, all of which contribute to strain hardening. This hardening effect arises because the growing density of dislocations and stacking faults makes it more difficult for new dislocations to move, thereby increasing the stress required for continued deformation. Eventually, the material reaches its yield point, beyond which local thinning, or necking, occurs. At this point, dislocation density is high, and interactions between dislocations and stacking faults may lead to local softening as dislocations start to rearrange or annihilate. Together, stacking faults and dislocations create a complex internal structure as shown in Fig 6 for different interatomic potentials and at their corresponding strain value where the plastic deformation starts. We noticed that EAM-2 forms several SF more than the other potentials due to the GSFE energy information in the potential, while EAM-2 and MEAM model the mechanical response in a similar manner.

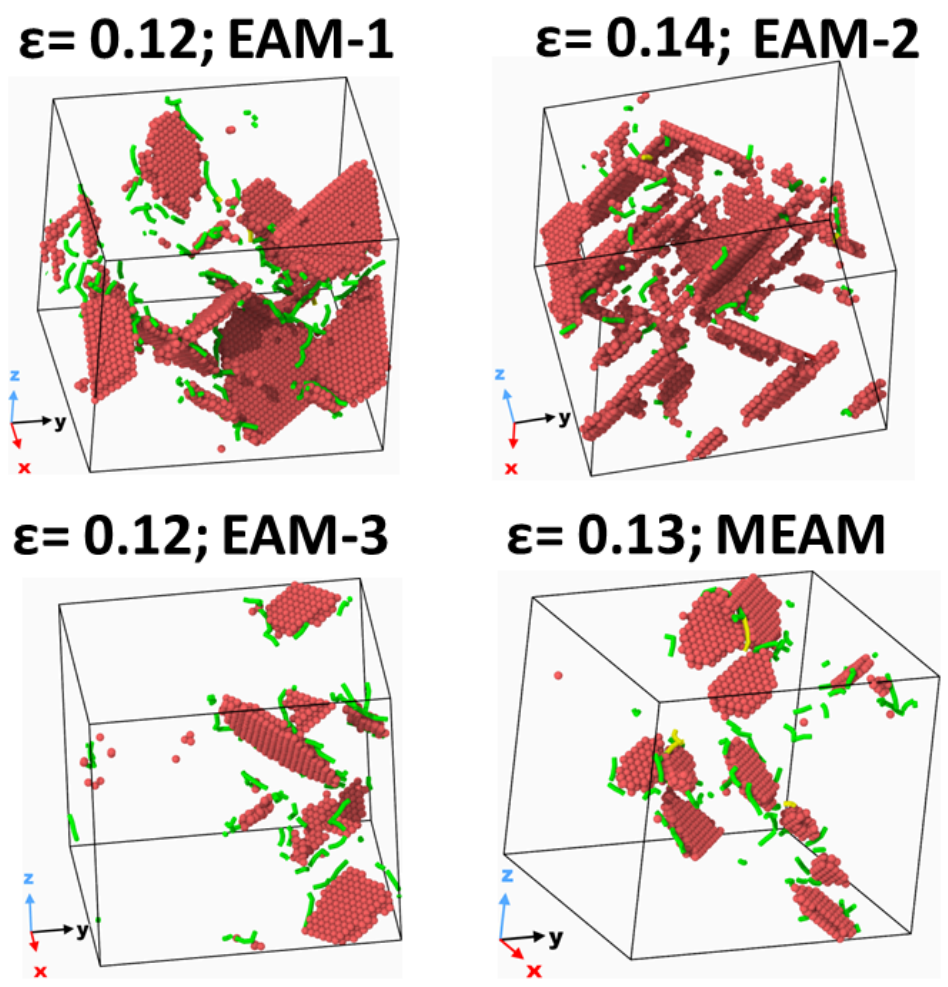

Fig 6. Stacking fault formation and dislocation nucleation at the point of plastic deformation initiation for [001] .

## 3.2 Nanoindentation test

During nanoindentation loading, the localized loading from the indenter tip initiates dislocation nucleation and multiplication at several stages during the loading, which are central to the material's plastic response. This mechanical response of the material can be analyzed from the load-displacement curves shown in Fig. 7 at different crystal orientations and by using different interatomic potentials. Initially, as the indenter makes contact with the surface, the material undergoes elastic deformation, where atomic bonds stretch but remain intact, allowing reversible displacement, where this elastic response can be fitted to a Hertz curve to compute the reduced elastic modulus [16-21]. As the indenter force increases, the stress beneath the tip exceeds the yield strength, triggering plastic deformation which is known as the pop-in event and identified as the deviation of the load curve with respect to the Hertz fitting curve [16,20,21]. Sharp increases in displacement, known as pop-in events, can occur in the load-displacement curve; these are typically associated with the sudden nucleation and movement of many dislocations, resulting in abrupt plastic deformation. Due to the high-stress concentration under the indenter, dislocations nucleate just beneath the surface. This process continues until the indenter tip is stopped. From our MD simulations, we observed the EAM potentials describe a similar mechanical response of the material under external mechanical load for the [100] orientation, in agreement with MEAM results [16]. However, for [101] and [111] the EAM-3 results oscillate more due to the GSFE energy values. Nevertheless, all interatomic potentials model the elastic part of the nanoindentation loading process in good agreement, where the reduced Young modulus will be the same value regardless of the chosen interatomic potential.

The Voigt-Reuss-Hill (VRH) approximation is a widely used method for estimating the effective elastic properties of materials based on their single-crystal elastic constants. It combines two theoretical bounds: the Voigt bound, which assumes uniform strain throughout the material, and the Reuss bound, which assumes uniform stress. The Voigt bound, representing a parallel distribution of crystals, provides an upper limit of the elastic modulus, while the Reuss bound, representing a series distribution of crystals, provides a lower limit. The VRH approximation takes the arithmetic mean of these bounds, offering a realistic estimate of macroscopic elastic properties [32, 33]. This approach is particularly relevant for FCC metals due to their cubic symmetry and relatively low anisotropy, which make them well-suited for averaging methods. FCC metals are characterized by three independent single-crystal elastic constants, $C_{11}$, $C_{12}$, and $C_{44}$, which describe their response to normal and shear stresses. Using these constants, the bulk modulus is calculated from the Voigt and Reuss expressions as $K_V = K_R = 1/3(C_{11} + 2C_{12})$, while the shear modulus is determined separately for each bound, with the Voigt approximation given by $G_V = 1/5(C_{11} - C_{12} + 3C_{44})$ and the Reuss approximation by $1/G_R = 5/2( 1/(C_{11} - C_{12}) + 2/C_{44})$. The final shear modulus is obtained as the arithmetic mean $G_{VRH} = (G_V + G_R)/2)$, while the Young's modulus and Poisson's ratio are derived from the bulk and shear moduli using standard isotropic elasticity relations. This methodology provides a robust framework for connecting the elastic properties of single crystals to the effective behavior of materials, making it particularly valuable for studying FCC metals and designing materials with tailored mechanical properties. Then, Young's modulus is calculated using the formula:

$$E_{VRH} = \frac{9K_{VRH}G_{VRH}}{3K_{VRH}+G_{VRH}} \quad (2),$$

where and  are the $K_{VRH}$ and $G_{VRH}$ averaged bulk and shear moduli, respectively. This methodology provides a robust framework for connecting the elastic properties of single crystals to the effective behavior of materials, making it particularly valuable for studying FCC metals and designing materials with tailored mechanical properties. In Tab 1, we present a comparison between the VRH elastic modulus and the reduced elastic modulus to show the nanomechanical response of the material during the early stages of nanoindentation for each interatomic potential. We noticed that the EAM-1 and MEAM are the potentials that can describe the nanoindentation mechanisms properly for the FCC Ni single crystal regardless of the crystal orientations, reaching a good agreement with reported experimental values for the [100] Ni orientation [34].

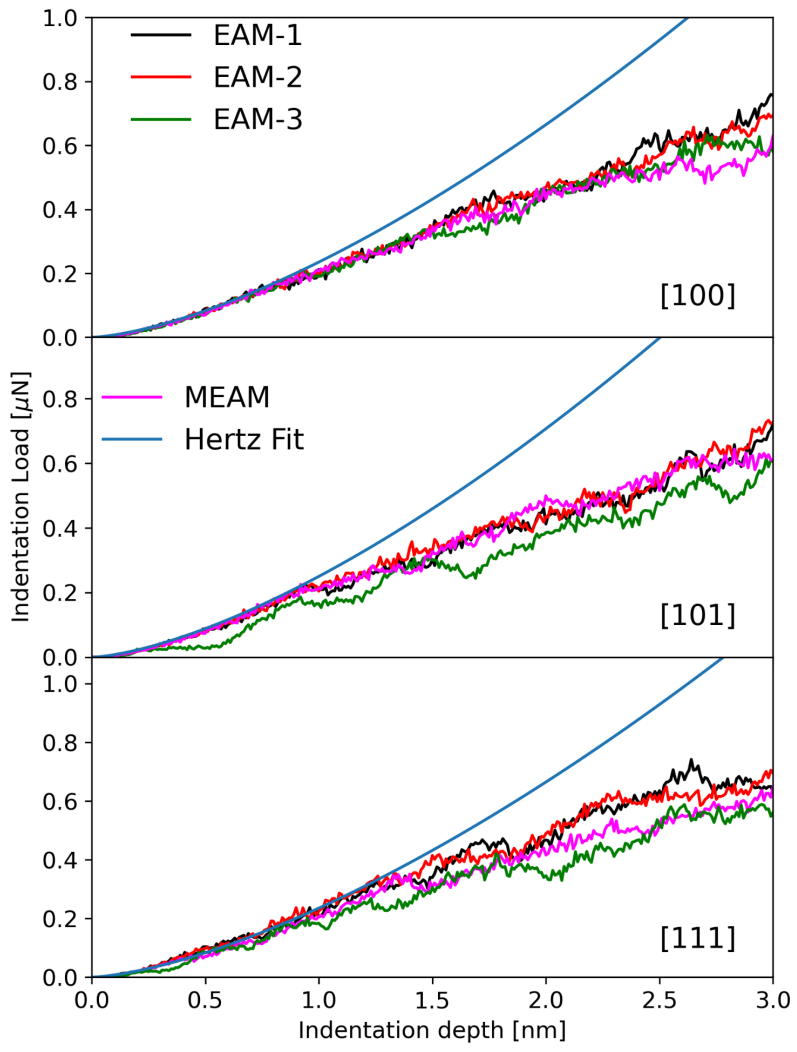

Fig. 7. Load displacement curves of nanoindentation loading at different crystal orientations and modeled by different interatomic potentials. A Hertz fitting curve is added to identify the elastic part of the process and the pop-in event.

Tab 1. Comparison between reduced elastic constant obtained by nanoindentation test and by Voigt-Reuss-Hill Approximation from the calculated $C_{ij}$ tensor, as well as experimental data for the [100] orientation from nanoindentation test [34]

| | EAM-1 | EAM-2 | EAM-3 | MEAM | Exp. [34] | EY |
|---|---|---|---|---|---|---|
| [001] | 205.7 | 218.5 | 194.5 | 209.1 | 204 | Reduced |
| | 221.3 | 218.5 | 223.5 | 239.9 | 227.25 | VRH |

In FCC Nickel, dislocations commonly nucleate as partial dislocations, in a similar way that in the compression simulation, each carrying a fraction of the full Burgers vector, which forms stacking faults and slip planes that help the material accommodate strain, as shown in Fig 8 at a indentation depth of 3nm. As the load increases, dislocations multiply and spread outwards, forming networks and allowing further plastic deformation like the formation of prismatic dislocations loops. In principle, dislocation multiplication mechanisms, such as those involving Frank–Read sources [27], generate additional dislocations that propagate on slip planes, enabling the material to “flow” under the indenter, as observed in our MD simulations for indentation depths beyond the pop-in event [18-21]. As dislocations move, they encounter obstacles like other dislocations, leading to pile-up, which in turn creates localized stress and contributes to work hardening [16,20,21]. This work hardening requires a higher load to maintain further indentation. With continued loading, a plastic zone develops under the indenter, characterized by a high density of dislocations and altered crystal structure, especially along slip planes in single crystals [16]. In Fig. 8, we present the visualization of the nanostructure of the indented samples at an indentation depth of 3nm by characterizing the HCP atoms to visualize the formation of prismatic dislocation loops, as well as the dislocation associated to these defects. In addition, we calculate the displacements to visualize the slip traces formed during the mechanical load, and the strain for the Ni samples at the [001] orientations. Although the formation of slip traces on the surface and the strain patterns in the Ni samples are modeled in good agreement among the interatomic potentials, the formation and evolution of prismatic dislocation loops is modeled in different manner. While EAM-1 and EAM-2 results already nucleated 4 prismatic dislocation loops, EAM-3 has managed to create 2, which can be attributed to the mechanical properties values for each interatomic potential. Finally, MEAM is able to model the nucleation of these defects in more detail due to the surface information and description of the atoms at nearest neighbors.

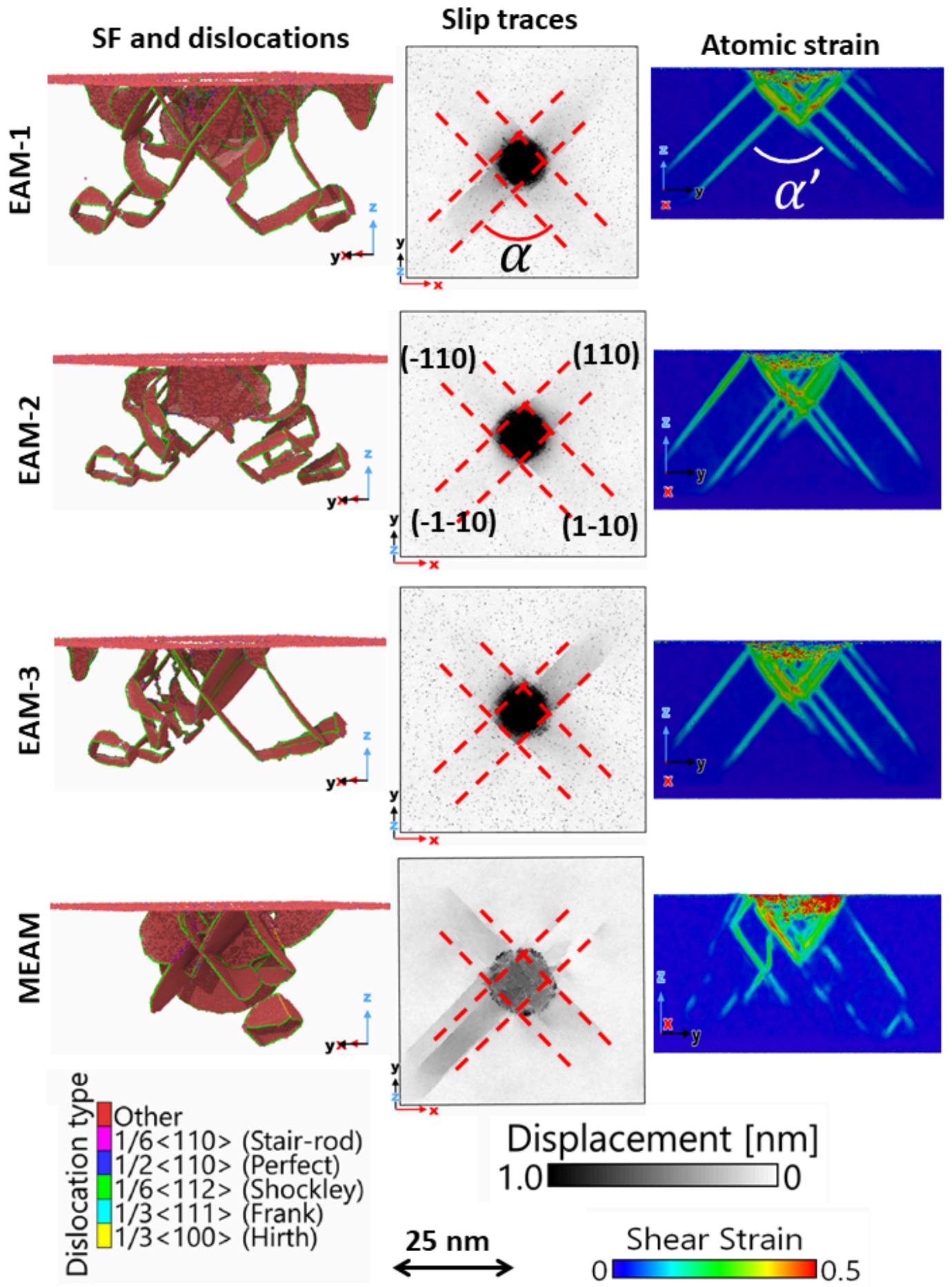

Fig. 8 Characterization of the indented Ni sample at [001] crystal orientation at an indentation depth of 3nm. Identifying the SF, prismatic dislocation loops, slip traces on the [101] and symmetric planes (highlighted by dashed lines), and atomic strain mapping patterns by different interatomic potentials.

Based on our simulations and the typical patterns of slip traces and strain mappings in FCC metals [16, 21], we observed that the angle, α, characterizing the 4-fold rosette pattern of slip traces, varies across different crystal orientations and being independent of the chosen interatomic potential. Specifically, this angle is larger for the [111] orientation compared to that observed for the [100] orientation in Fig. 8, and it is smaller for the [011] orientation. Additionally, another angle, α', can be identified between the slip planes in the strain mapping. For the [011] orientation, α' is zero, indicating that loop propagation is parallel to the orientation planes, while for the [111] orientation, the angle α' is reduced. These findings highlight orientation-dependent differences in slip behavior and strain distribution.

Nanoindentation and uniaxial compression can reveal different deformation mechanisms due to the difference in scale and the nature of loading. For example, during uniaxial compression, dislocation motion and grain boundary interactions throughout the sample affect the response, while nanoindentation often initiates localized dislocations, stacking faults, or phase transformations, especially in the material beneath the indenter tip, as we have discussed. In order to provide more information about the advantages and limitations of the interatomic potentials to properly model nanoindentation mechanisms. In Fig. 9, we present the stress–strain curves derived from uniaxial compression and nanoindentation simulations, following the protocol described in Ref. 26. For comparability, the strain values from nanoindentation simulations were scaled to match those of the compression simulations. We observe good alignment between the two mechanical tests, particularly in the description of the elastic-to-plastic transition, where the computational models and interatomic potentials capture the material behavior effectively. The yield points are consistent across both tests, further supporting the reliability of the simulations for EAM-1 and MEAM cases. An elastic reference curve, $\sigma = E\epsilon$, is included to clearly delineate the elastic and elastic-plastic regions under load. This agreement between the two methods suggests that nanoindentation results can serve as a localized representation of bulk mechanical properties and that the selected modeling approach is robust for studying similar materials under varied loading conditions.

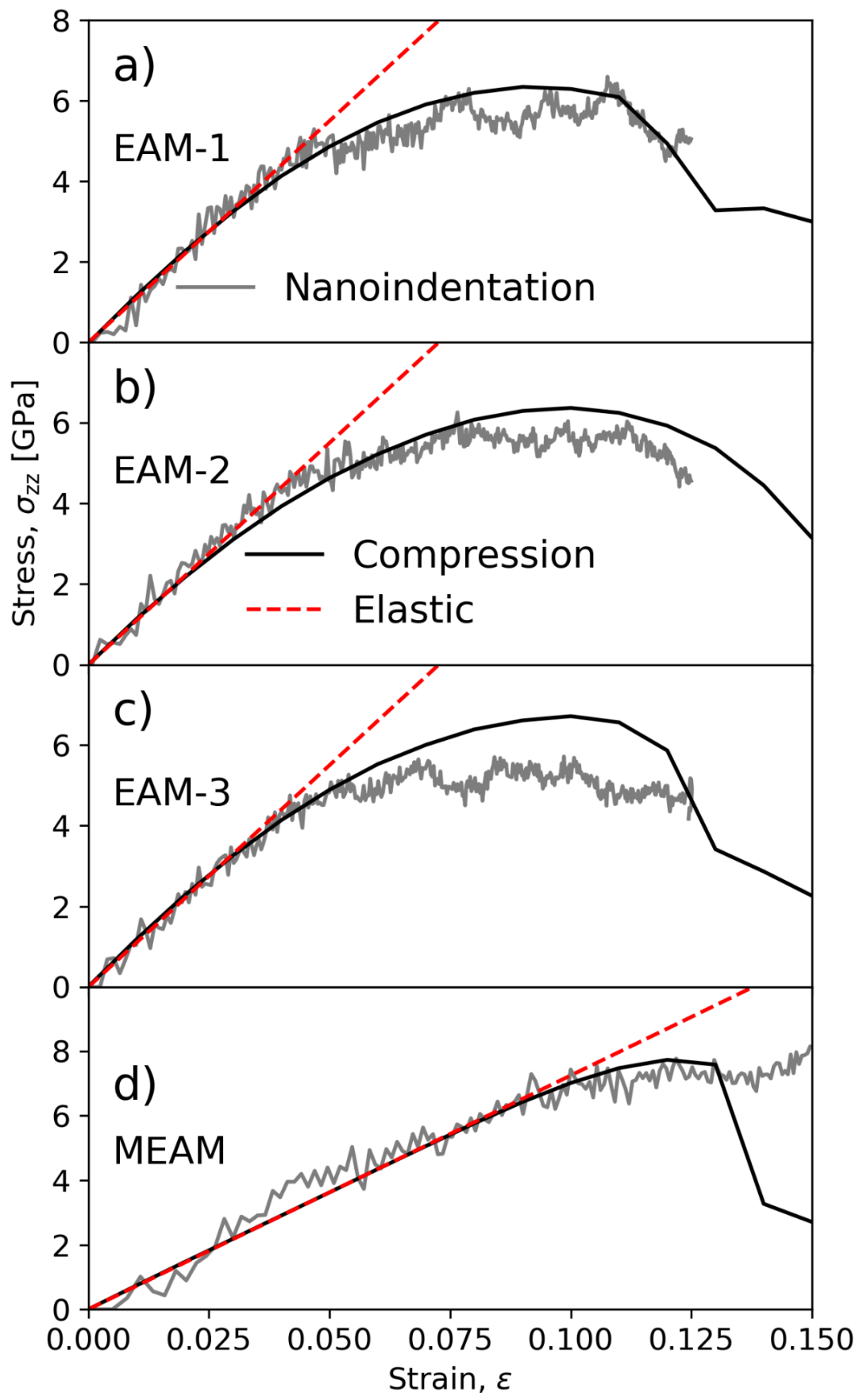

Fig 9. Stress-strain curves for compression and nanoindentation simulations on the [001] Ni sample. We compare results for different interatomic potentials, noticing that MEAM is able to provide a good description of the plastic deformation of the material for both mechanical tests.

## 4. Concluding remarks

In summary, we have conducted molecular dynamics simulations to investigate the response of FCC Ni under uniaxial compression and nanoindentation using traditional interatomic potentials, highlighting the limitations and strengths of EAM- and MEAM-based approaches. For each potential, we calculated the generalized stacking fault energy (GSFE) to examine the dissociated slip paths responsible for stacking fault formation and partial Shockley dislocations during mechanical loading. Our compression simulations reveal that EAM potentials tend to describe the material's plastic deformation after the yield point, though some EAM potentials have been modified to approximate dislocation formation, they generally lack detail in this area. In contrast, MEAM potentials offer a more accurate description of both the elastic regime and the transition to plastic deformation, effectively modeling phenomena such as twinning and the formation of Shockley and Hirth dislocations, while requiring more computational resources. Similar trends were observed in nanoindentation simulations: load-displacement curves showed no notable differences between EAM and MEAM results, the nucleation of prismatic dislocation loops and the development of the dislocation network beneath the indenter tip were better captured by the MEAM potential. This suggests that MEAM can more reliably model both uniaxial compression and nanoindentation for pure FCC Ni. Consequently, our study presents a practical methodology for selecting empirical interatomic potentials to perform mechanical tests by MD simulations, our results indicate that while NiFe alloys can be effectively modeled by both EAM-1 [21] and MEAM [24] potentials modeling properly FCC Ni samples, more complex alloys are likely best represented by MEAM. The outcomes of this study establish a benchmark for pure FCC Ni and provide a foundation for extending simulations to more complex systems.

## Acknowledgments

Research was funded through the European Union Horizon 2020 research and innovation program under Grant Agreement No. 857470 and from the European Regional Development Fund under the program of the Foundation for Polish Science International Research Agenda PLUS, Grant No. MAB PLUS/2018/8, and the initiative of the Ministry of Science and Higher Education “Support for the activities of Centers of Excellence established in Poland under the Horizon 2020 program” under Agreement No. MEiN/2023/DIR/3795. We gratefully acknowledge Polish high-performance computing infrastructure PLGrid (HPC Center: ACK Cyfronet AGH) for providing computer facilities and support within computational Grant No. PLG/2024/017084. K.C. acknowledges support by Dean's grant 16.16.110.663- task 13 AGH University of Krakow. The research project supported by the program “Excellence initiative—research university” for the AGH University of Krakow.